\documentclass[twoside,12pt]{report}
\usepackage{graphicx}
\usepackage{amssymb, amsmath, amsxtra}
\usepackage{epsfig,subfigure}

\usepackage{latexsym}
\usepackage{bm}

 \usepackage{ntheorem}
           \theoremstyle{plain}
                      {\theorembodyfont{\rmfamily}
                      \theoremseparator{.}

           \theoremstyle{plain}
           
           \theoremstyle{plain} 
           \theoremstyle{plain}

           \theoremstyle{plain}
           
            }

\begin{document}

\begin{center}
{\Large \textbf{Evolution equations of the second and third order with Lie-B\"{a}cklund symmetries}}\\[2ex]
% {\Large \textbf{Lie-B\"{a}cklund symmetries of the Krichever-Novikov equation}}\\[2ex]
 Nail H. Ibragimov\\
 Department of Mathematics and Science, Blekinge Institute
 of Technology,\\ 371 79 Karlskrona, Sweden
 \end{center}

 \section*{Preface}
 %\noindent
 %\textbf{Abstract.}
 Chapter 4 of my book \cite{ibr83} is dedicated to
 Lie-B\"{a}cklund group analysis of various types of differential equations.
 The Russian edition of this book
 had an \textit{Addendum} (pages 262-266) containing a summary of
 new results (obtained by the end of 1982) on classification of evolution equations
 of the second and third order possessing Lie-B\"{a}cklund symmetries.
 The \textit{Addendum} was not included in the English translation of the book by technical reasons.
 I present here the missing translation. In order to make the text  self-contained, I
 have added in the translation
 the equations from the main body of the book used in the \textit{Addendum}.

 The $m$th-order evolution equation with one spatial variable
 $x$ is written
 \begin{equation}
 \tag{E}
 \label{E}
 u_t = F(x, u, u_1,..., u_m), \quad m \geq 2,
 \end{equation}
where $u_s, \ s = 1, \ldots, m,$ is the partial derivative of order
$s$ of $u$ with respect to $x.$\linebreak The expression
 $
 F_m = \partial F(x, u, u_1,..., u_m)/\partial u_m
 $
 is called the \textit{separant} of Eq. (\ref{E}).  \\[7ex]
 %\noindent
 %\textit{Keywords}: Evolution equations, Lie-B\"{a}cklund symmetries, Calogero-Degasperis equation,
 %Krichever-Novikov equation.\\[20ex]
%  \noindent
% MSC: 70S10, 35C99, 35G20\\
% \noindent
% PACS: 02.30.Jr, 11.15.-j, 02.20.Tw\\[2ex]
{\Large \bf Contents}\\[.2ex]
 \contentsline {section}{\numberline {I}}{2}
\contentsline {section}{\numberline {II}}{4} \contentsline
{section}{\numberline {III}}{6} \contentsline {section}{\numberline
{IV}}{7} \contentsline {chapter}{Bibliography}{7}

 \hfill

 \begin{table}[b]
  \begin{tabular}{l}\hline
 \copyright ~2011 N.H. Ibragimov.\\ Translation from \textit{N.H. Ibragimov, Transformation groups applied to
 mathematical}\\
  \textit{physics}, Nauka, Moscow, 1983, Addendum, pp. 262-266.\\
  \end{tabular}
 \end{table}

%\tableofcontents
%  \null
  \newpage

% Supplement

 \section{}
\label{subsection A}

% The supplement provides results of group classification of
% evolutionary equations obtained by the end of 1982.

The problem on classification of evolution equations having
 Lie-B\"{a}cklund symmetries has been solved in
\cite{ibr83}, Section 20, for
 the semi-linear second-order equations of the form
 $$
u_t=\varphi(u)u_2+\psi (u,u_1)
 $$
 and for the third-order
equations with a constant separant having the form
 $$
  u_t = u_3 + \varphi(u,u_1).
 $$
 This problem has been recently studied in more general cases.

S.I.~Svinolupov and V.V.~Sokolov  \cite{SS82} considered the
second-order evolution equations of the form
 $$
u_t= F(u, u_1, u_2)
 $$
and extended the analysis of necessary conditions for existence of
the recursion operator (see Eqs. (20.8), (20.9), (20.9$'$) in
\cite{ibr83}) by deriving three additional necessary conditions.
They obtained the following forms of the function $F:$
 $$
\frac{u_2}{u^2_1}-\frac{a^{''}}{a'} + bu_1, \quad \frac{u_2}{u^2_1}
+ \frac{1}{u_1}+ bu_1 +c,
$$
 $$
 \frac{u_2}{(u_1+1)^2} - \frac{b'-k^2}{b+k}\, \frac{1}{u_1+1} +
\frac{b^2-b'}{b+k}(u_1+1) + 2 \frac{b'+kb}{b+k}\,,
$$
 $$
 \frac{u_2}{(u_1+1)^2} + \frac{a''}{a'} \, \frac{1}{u_1+1} +
  \left( \frac{a''}{a'} +ka \right)u_1-\frac{a''}{a'},
 $$
 $$
 \frac{u_2+ a'u_1}{(u_1+1)^2} + \frac{aa''}{a'(u_1+a)} -
 \left( \frac{a''}{a'} -\frac{a'}{a^2}+ \frac{k}{a^2} \right) u_1,
$$
where $k=$const., $a,$ $b,$ $c$ are arbitrary functions of $u$ and
$a(u)$ is the density of the conservation law of the corresponding
equation. These functions describe all possibilities of the cases
(20.9) and (20.9$'$) from \cite{ibr83}. An additional analysis is
necessary for the case (20.8).

The condition (19.37) from \cite{ibr83} yields that the third-order
equation with the constant separant $F_3$ possessing a non-trivial
Lie-B\"acklund algebra has the form
 \begin{equation}
 \label{sup1}
 u_t=u_3+ a(u, u_1)  u^2_2 +b (u, u_1)u_2 +c (u,u_1).
 \end{equation}
The equations (\ref{sup1}) with $a=b=0$ are classified in
\cite{ibr83}, Section 20.2. Furthermore,
 the following two integrable equations of the form (\ref{sup1}) were known aside
equations reducible to the case $a=b=0$  by a simple transformation.
The equation \cite{CD79}
 \begin{equation}
 \label{sup2}
 u_t=u_3-\frac32\, \frac{u_1}{u^2_1+\alpha} \,u^2_2 -\frac32\,
 \frac{\alpha'u_1}{ u^2_1+\alpha} u_2-\frac38\alpha'^2 \,
\frac{u_1}{u^2_1+\alpha}+\frac12 \alpha'' u_1,
 \end{equation}
   where $\alpha=\sum\limits^4_{i=1}k_i (u+k)^i$ is an arbitrary fourth degree polynomial  in $u$  with real coefficients,
   and the equation \cite{KN80}
  \begin{equation}
 \label{sup3}
 u_t=u_3-\frac32 \,\frac{u^2_2}{u_1} -\frac32 \wp(u) u^3_1 +\frac{k}{u_1}, \quad k\neq 0,
\end{equation}
 where $k=$const., $\wp(u)$ is the Weierstrass elliptic function,
 $$
\wp'^2=4 \wp^3-g_2 \wp-g_3 \equiv 4(\wp-e_1)(\wp-e_2)(\wp-e_3),
$$
 $$
 e_1+e_2+e_3=0.
$$

 Upon enumerating \cite{SS82a} all possible equations (\ref{sup1}) with a nontrivial
 algebra (assuming that  elements of the algebra are independent of
$t,x$), it became clear that Eqs. (\ref{sup2}), (\ref{sup3})
 are exceptional, and the remaining equations (\ref{sup1}) are reducible to the KdV
equation or to a linear equation by means of rather simple
transformations. Recently, S.V.~Khabirov (see \ref{subsection B})
and independently  S.I.~Svinolupov, V.V.~Sokolov (see
\ref{subsection C}), have found transformations of the form
  \begin{equation}
 \label{sup4}
  v=\Phi(u, u_1, \ldots, u_n),
\end{equation}
connecting (\ref{sup2}) with the KdV equation, and proved that there
are no such transformations for Eq. (\ref{sup3}), except for
degenerate cases. Thus, classification of third-order equations with
the constant separant possessing a nontrivial Lie-B\"{a}cklund
algebra results in the linear equation, the KdV equation and the
\textit{Krichever-Novikov equation} (\ref{sup3}). This solves the
problem of classification of semi-linear equations
 $$
 u_t=a(u)u_3+\varphi (u, u_1, u_2)
 $$
 as well. Indeed, if $a'\neq0$ one can assume that
$a(u)=u^3$ (using the substitution $a(u)=v^3$), and such equations
with a nontrivial algebra are reduced to the case of the constant
separant by the substitution (20.44) from \cite{ibr83}.

 For the equation of the form (\ref{sup3}) with $k=0$ (in this case one can take an arbitrary function of $u$ instead of
 $\wp(u)$)
one can readily obtain the recursion operator, because a
substitution $w =\phi(u)$ reduces it to Eq. (20.43) from
\cite{ibr83}:
$$
 w_t=w_3-\frac32 w^{-1}_1 w^2_2.
  $$
 The latter equation has the
recursion operator
 $$
   L= D^2 -2 \frac{w_2}{w_1}D +w_1 D^{-1} \cdot \left( \frac{w_3}{w^2_1} -\frac{w^2_2}{w^3_1} \right) D
$$
 which is obtained by the formulae (19.46), (19.50$'$), and (20.42) from \cite{ibr83}.

 \section{}
\label{subsection B}

(S.V.~Khabirov). Equivalence transformations of the form
(\ref{sup4}) are studied for equations
 \begin{equation}
 \label{sup5}
  u_t=u_3+f(u,  u_1, u_2),
 \end{equation}
 \begin{equation}
 \label{sup6}
  v_t=v_3+h (v,  v_1, v_2)
 \end{equation}
 without assuming that the equations (\ref{sup5}) and (\ref{sup6}) admit a nontrivial algebra.
  All equations are considered up to pointwise changes of variables, so that
transformations (\ref{sup4}) of the order $n\geq1$ are discussed.

 It turns out that existence of equivalence transformations imposes strict restrictions on functions $f$ and $h.$ For example, if the KdV equation
($h=vv_1$) is taken as (\ref{sup6}), it appears that (\ref{sup5})
has the form (\ref{sup1}) and the transformation (\ref{sup4}) has
the order $n\leq3.$ Furthermore, one can enumerate the equations
(\ref{sup1}) reducible to the KdV equation by transformations of the
first, second, and third order and to find the transformations
(\ref{sup4}) themselves. In particular, (\ref{sup2}) is connected
with the KdV equation
 \begin{equation}
 \label{sup7}
  v_t=v_3+vv_1
 \end{equation}
by the  third-order transformation
 $$
 v=pu_3+qu^2_2+ru_2+s,
$$
  where
 $$
 p=3\frac{z}{u_1}, \quad q= -\frac{3}{2\alpha} (1-z^2) (1+2z),
$$
 $$
 r=\frac{6(1-z)}{u+k} +\alpha\,'q,
$$
 $$
 s=\frac{\alpha''}{2} +6\frac{\alpha}{(u+k)^2} +3z \left( \frac{\alpha''}{2}-\frac{\alpha}{u+k} \right)+\frac{\alpha'^2}{4}q,
$$
 $$
 z=\pm\frac{u_1}{\sqrt{u^2_1+\alpha}}\,\cdot
$$

 The following equations are connected with (\ref{sup7}) by second-order transformations
(every equation is followed by the corresponding transformation):
 \begin{align}
& u_t=u_3-\frac34 \frac{u_2^2}{u_1}-\frac13 u^2_1 -\frac23
ku^{3/2}_1, \quad &v=\frac{u_2}{\sqrt{u_1}} -\frac23 u_1
+k\sqrt{u_1};\notag\\
 & u_t=u_3-\frac{1}{18} u^3_1 +\frac12 ku^2_1,  &v=u_2-\frac16 u^2_1
 +ku_1;\quad\quad\notag\\
  &  u_t=u_3+3\frac{a'}{a} u_1u_2 + \left(  \frac{a''}{a} -\frac{a^2}{18}  \right)u^3_1, \quad &v=au_2+ \left( a'-\frac{a^2}{6} \right) u^2_1.\notag
 \end{align}
 Here $k$ is an arbitrary constant and $a=a(u)$ is an arbitrary function.

 The Krichever-Novikov equation (\ref{sup3}) is not connected with any equation (\ref{sup6}) by non-point transformations
 of the form (\ref{sup4}) in the general case.
 However, there are exceptions when it can be reduced to the KdV equation.
Namely, Eq. (\ref{sup3}), written here with $k=6,$ is connected with
(\ref{sup7}) by the transformation
\begin{equation}
 \label{sup8}
  v=3\left( \frac{u_3}{u_1}-\frac32 \frac{u^2_2}{u^2_1} + 4 \varepsilon \frac{u_2}{u_1^2}-\frac32 \wp u^2_1-\frac{2}{u^2_1}  \right), \quad
  \varepsilon=\pm1,
 \end{equation}
 if $\wp=$const, and by the transformation
\begin{equation}
 \label{sup9}
  v=-3 \left( \frac{u_3}{u_1}- \frac12 \frac{u^2_2}{u^2_1} +\varepsilon(u)u_2 +\varepsilon'(u)u^2_1 +\frac32 \wp(u)u^2_1 +\frac{2}{u^2_1} \right)
 \end{equation}
if
\begin{equation}
\tag{\ref{sup9}a}
 \label{sup9a}
   \wp=\frac{1}{u^2}, \qquad \qquad \qquad {\rm then} \quad \varepsilon=\frac{2}{u},
 \end{equation}
or
\begin{equation}
\tag{\ref{sup9}b}
 \label{sup9b}
   \wp=\frac{\alpha^2}{4} \left( -\frac23+\tan^2\frac{au}{2} \right),   \quad {\rm then}\quad \varepsilon=\alpha \tan \frac{au}{2},
 \end{equation}
 or
 \begin{equation}
 \tag{\ref{sup9}c}
 \label{sup9c}
   \wp=\frac{\alpha^2}{4}\left( \frac23+\tan^2\frac{au}{2}   \right), \qquad \ {\rm then}\quad \varepsilon=\alpha\tanh\frac{au}{2},
 \end{equation}
 where $\alpha=$const.

 The Lie-B\"acklund algebra for Eq. (\ref{sup3}) with the arbitrary Weierstrass function is nontrivial and contains the following element of the fifth order:
 $$
 u_5-5\frac{u_2u_4}{u_1}-\frac52 \frac{u^2_3}{u_1} +\left( \frac{25}{2} \frac{u^2_2}{u^2_1}
 -\frac52 \frac{k}{u^2_1} -\frac{15}{2} \wp u^2_1  \right) u_3 -\frac{45}{8}\frac{u^4_2}{u^3_1} +\frac{25}{4} k \frac{u^2_2}{u^3_1} +
 $$
 $$
 +\frac{15}{4} \wp u_1 u^2_2 -\frac{15}{2} \wp' u^3_1 u_2 -\frac32\wp'' u^5_1 +\frac{27}{8} \wp^2 u^5_1 -\frac58 \frac{k^2}{u^3_1}+\frac54 k\wp u_1.
 $$

 \section{}
 \label{subsection C}

 (S.I.~Svinolupov, V.V.~Sokolov). Eq. (\ref{sup2}) can be connected with Eq. (20.32) from \cite{ibr83} by the transformation (\ref{sup4}) of the first order.
To this end, one should first reduce (\ref{sup2}) to the form
\begin{equation}
 \tag{\ref{sup2}$'$}
 \label{sup2'}
    u_t=u_3-\frac32 \frac{u_1}{u^2_1+1} u^2_2 -\frac32 \wp (u) (u^3_1+u_1)
 \end{equation}
 by a pointwise substitution and then perform the transformation
 $$
 v=2\ln \left(u_1+\sqrt{u^2_1+1}\right)+\ln\psi(u)
$$
with the function $\psi(u)$ defined by the equation
 $$
 A\psi^2+\left( \frac32 \wp (u) +C \right)\psi+B=0.
$$
 If the coefficients $A, B, C$ are expressed via the irrational invariants of the function $\wp(u)$
by the formulae
 $$
 AB=\frac{9}{64}(e^2_1-4e_2e_3), \quad C=\frac{3}{4}\, e_1,
$$
 the above transformation reduces (\ref{sup2'}) to
 $$
 v_t=v_3-\frac18 v^3_1 +(Ae^v+Be^{-v}+C)v_1.
$$
 Then, one can construct a third-order transformation connecting (\ref{sup2'}) with (\ref{sup7}) according to Lemma 20.2.2 from \cite{ibr83}.

 The following chain of transformations  is suggested for Krichever-Novikov equation (\ref{sup3}).
 It allows one to investigate the equation in question from various viewpoints.
The substitution $v=\wp\left(\frac{u}{2}\right)$ maps (\ref{sup3})
to the form
\begin{equation}
\tag{\ref{sup3}$'$}
 \label{sup3'}
 v_t=v_3-\frac32 \frac{v^2_2}{v_1} +\frac{av^3+bv+c}{v_1}
  \end{equation}
with the constants $a, b, c.$ This equation is equivalent to
(\ref{sup7}) when the cubic  polynomial $av^3+bv+c$ has multiple
zeroes. In the general case, it is reduced by the transformation
 $$
 w=-3\frac{v_3}{v_1} +\frac32 \frac{v^2_2}{v^2_1} -\frac{av^3+bv+c}{v^2_1}
$$
to the system
 $$
 v_t=-2v_3-w v_1, \quad w_t=w_3+ww_1-12 av_1
$$
 with the known $(L, A)$-pair \cite{DS81}.

 \section{}
  \label{subsection D}

 The equivalence problems discussed above can be considered in a more general framework
 by replacing the transformations (\ref{sup4}) solved for $v$ with
 a more general transformation between the
differentiable variables $u, v $ given by the differential equation
 \begin{equation}
 \label{sup10}
 \Phi (u, u_1, \ldots, u_n; v, v_1, \ldots, v_n)=0.
 \end{equation}
 %instead of the transformations (\ref{sup4}) solved with respect to $v.$
 Let $[\Phi]$ be a differential manifold in the space of variables
$(u, v; u_1, v_1; \ldots)$ given by Eq. (\ref{sup10}). The
evolutionary equations
\begin{equation}
 \label{sup11}
u_t=F(u, u_1,\ldots, u_m),
\end{equation}
\begin{equation}
 \label{sup12}
 v_t=H(v, v_1, \ldots, v_m)
\end{equation}
will be considered as the Lie-B\"acklund equations (see Remark
17.1.1 in \cite{ibr83})\linebreak determining the group $G$ with the
canonical generator
 $$
 X=F\frac{\partial }{\partial u} +H \frac{\partial }{\partial v} +\ldots.
$$
 The equations (\ref{sup11}) and (\ref{sup12}) are equivalent if there is a manifold $[\Phi]$
invariant with respect to the group $G.$ The equivalence
transformation  between the equations (\ref{sup11}) and
(\ref{sup12}) is given by the differential equation
 (\ref{sup10}). Thus, the problem of equivalence of evolutionary equations is reduced to investigation of the
invariance test
 $$
 X\Phi\vert_{[\Phi]}=0.
 $$
 The similar generalization of
 the transformation (19.44) from \cite{ibr83},
 $$
 y = \varphi (x,u,u_1 ,...,u_n ),\quad  v= \Phi (x,u,u_1,...,u_n),
 $$
including a change of the variable $x$, leads to the general
\textit{B\"acklund transformation} for evolutionary equations.

% \begin{thebibliography}{99}

 \end{document}